\documentclass[preprint,12pt,times,review,authoryear]{elsarticle}

\journal{EPSL}

\usepackage{graphicx}
\usepackage{epstopdf}
\usepackage{txfonts}
\usepackage{hyperref}
\usepackage{booktabs}
\usepackage{rotating}
\usepackage{longtable}
\usepackage[font=small,labelfont=bf]{caption}
\usepackage{lineno}
\usepackage{setspace}
\usepackage{amsmath}
\usepackage{wasysym}
\usepackage{mhchem}

\providecommand{\E}[1]{\ensuremath{10^{#1}}}
\providecommand{\isotope}[2]{\ce{^{#2}#1}}
\providecommand{\HeThree}{\ce{^3He}}
\providecommand{\HeFour}{\ce{^4He}}
\providecommand{\NeTwenty}{\ce{^{20}Ne}}
\providecommand{\NeTwentyTwo}{\ce{^{22}Ne}}
\providecommand{\ArTSix}{\ce{^{36}Ar}}
\providecommand{\ArFourty}{\ce{^{40}Ar}}
\providecommand{\KrESix}{\ce{^{86}Kr}}
\providecommand{\XeHTSix}{\ce{^{136}Xe}}
\providecommand{\tritium}{\ce{^3H}}
\providecommand{\Otwo}{\ce{O_2}}
\providecommand{\COtwo}{\ce{CO_2}}
\providecommand{\methane}{\ce{CH_4}}
\providecommand{\Htwo}{\ce{H_2}}
\providecommand{\Ntwo}{\ce{N_2}}
\providecommand{\NtwoO}{\ce{N_2O}}
\providecommand{\SOtwo}{\ce{SO_2}}
\providecommand{\ozone}{\ce{O_3}}
\providecommand{\water}{\ce{H_2O}}
\providecommand{\Edot}[1]{\ensuremath{\times \E{#1}}}
\providecommand{\RHe}{\HeThree/\HeFour}
\providecommand{\RHeNe}{\HeFour/\NeTwenty}
\providecommand{\RNe}{\NeTwenty/\NeTwentyTwo}
\providecommand{\RAr}{\ArFourty/\ArTSix}

\providecommand{\dwn}[1]{\ensuremath{_{\text{#1}}}}
\providecommand{\cubmSTP}{\ensuremath{\text{m}^{\text{3}}\dwn{STP}}}
\providecommand{\ccSTP}{\ensuremath{\text{cm}^{\text{3}}\dwn{STP}}}
\renewcommand{\deg}{\ensuremath{{}^\circ}}
\providecommand{\ppmV}{\ensuremath{\text{ppm$_{\rm vol}$}}}
\providecommand{\permille}{\ensuremath{\permil}}

\providecommand{\seclabel}[1]{\label{sec:#1}}
\providecommand{\secn}[1]{Sec.~\ref{sec:#1}}
\providecommand{\tablabel}[1]{\label{tab:#1}}
\providecommand{\tabl}[1]{Tab.~\ref{tab:#1}}
\providecommand{\figlabel}[1]{\label{fig:#1}}
\providecommand{\figr}[1]{Fig.~\ref{fig:#1}}

\begin{document}

\begin{frontmatter}


\title{Concentrations and isotope ratios of helium and other noble gases in the Earth's atmosphere during 1978--2011}


\author[eawag]{Matthias S.\ Brennwald\corref{corr}}
\ead{matthias.brennwald@eawag.ch}

\author[eawag,igp]{Nadia Vogel}
\ead{nadia.vogel@eawag.ch}

\author[eawag]{Simon Figura}
\ead{simon.figura@eawag.ch}

\author[empa]{Martin K.\ Vollmer}
\ead{martin.vollmer@empa.ch}

\author[csiro]{Ray Langenfelds}
\ead{ray.langenfelds@csiro.au}

\author[csiro]{L.\ Paul Steele}
\ead{paul.steele@csiro.au}

\author[igp]{Colin Maden}
\ead{maden@erdw.ethz.ch}

\author[eawag,igp]{Rolf Kipfer}
\ead{rolf.kipfer@eawag.ch}

\cortext[corr]{Corresponding author}
\address[eawag]{Eawag, Swiss Federal Institute of Aquatic Science and Technology, Dept.~of Water Resources and Drinking Water, 8600 D\"ubendorf, Switzerland}
\address[igp]{ETH, Swiss Federal Institute of Technology Zurich, Institute of Geochemistry and Petrology, 8092 Z\"urich, Switzerland}
\address[empa]{Empa, Swiss Federal Laboratories for Materials Science and Technology, Laboratory for Air Pollution and Environmental Technology, 8600 D\"ubendorf, Switzerland}
\address[csiro]{Centre for Australian Weather and Climate Research/CSIRO Marine and Atmospheric Research, Aspendale, Victoria, Australia}


\begin{abstract}
The evolution of the atmospheric noble gas composition during the past few decades has hardly been studied because, in contrast to many other atmospheric gases, systematic time-series measurements have not been available. Based on theoretical considerations, the atmospheric noble gas isotope composition is assumed to be stable on time scales of up to about \E{6} years, with the potential exception of anthropogenic changes predicted for the He concentration and the \RHe\ ratio. However, experimental assessments of the predicted changes in the atmospheric He isotope composition are controversial. To empirically test these assumptions and predictions, we analysed the noble gas isotope composition in samples of the Cape Grim Air Archive, a well-defined archive of marine boundary layer air in the southern hemisphere. The resulting time series of the \NeTwenty, \ArFourty, \KrESix\ and \XeHTSix\ concentrations and \RNe\ and \RAr\ ratios during 1978--2011 demonstrate the stability of the atmospheric Ne, Ar, Kr and Xe composition during this time interval. The He isotope data provide strong evidence for a decrease in the \RHe\ during the same time interval at a mean rate of 0.23--0.30\permille\ per year. This result is consistent with most model predictions of the rate of decrease in the atmospheric \RHe\ ratio associated with mining and burning of fossil fuels.
\end{abstract}
 


\end{frontmatter}

\clearpage


\linenumbers

\section{Introduction}
The evolution of numerous trace gases in the terrestrial atmosphere (e.g., \COtwo, \methane, \Htwo, \NtwoO, CO, \ozone, \SOtwo, nitrogen oxides, halocarbons, etc.) has been monitored and studied very comprehensively during the past few decades \citep{Forster:2007}. In contrast, no such systematic research has been conducted on the atmospheric noble gases (He, Ne, Ar, Kr and Xe), although they range among the 10 most abundant trace gases in the dry atmosphere.\par

The atmospheric composition with respect to Ne, Ar, Kr and Xe isotopes in the atmosphere is generally assumed to be stable on a time scale of about \E{6} years, because there are no known sources or sinks that might modify the atmospheric inventory of these gases on this time scale \citep{Ozima:2002}. In contrast, the concentration of He isotopes in the atmosphere is governed by the dynamic balance of their sources and sinks, i.e., the accumulation of terrigenic He by outgassing of the solid earth, the contribution of extraterrestrial He, and the He loss into space \citep[e.g.,][]{Kockarts:1973,Lupton:1983,Mamyrin:1984}. The mean residence time of He in the atmosphere is approximately \E{6} years \citep{Lupton:1983,Mamyrin:1984}. The atmospheric He isotope concentrations resulting from the dynamic balance in the natural geological sources and sinks may be variable on time scales of about \E{5}\,years or more, but are assumed to be stable on a time scale of a few millennia \citep[e.g.,][]{Lupton:1983,Oliver:1984,Mamyrin:1984,Pierson-Wickmann:2001}.\par

The natural He balance is possibly being disturbed by the release of He isotopes into the atmosphere as a result of  anthropogenic activities. In particular, an increase of the He concentration of a few permille during the past few decades has been predicted as a result of the mining and burning of fossil fuels \citep{Oliver:1984}. Fossil fuels often contain large amounts of terrigenic He, which is released into the atmosphere during processing and combustion of the fuel. However, detection of the predicted change in the atmospheric He concentration is challenging, because the precisions of currently available methods for determination of He concentrations in air are too low. In addition, chemical transformations of reactive gases (e.g., \Otwo , \COtwo, \water, etc.) in an archived air sample might result in a modification of the air matrix and thereby also in the He mole fraction.\par

An alternative approach to study the evolution of the atmospheric He composition is, in analogy to the Suess effect for the carbon isotopic composition of atmospheric \COtwo\ \citep[e.g.,][]{Tans:1979}, to analyse the \RHe\ ratio in archived air. The \RHe\ ratio in fossil fuels (\RHe$\approx$\E{-8}--\E{-7}) is commonly one to two orders of magnitude lower than the atmospheric ratio (\RHe=1.40\Edot{-6}). The predicted decrease in the \RHe\ ratio associated with the mining and burning of fossil fuels should be detectable with currently available analytical methods, and is unaffected by chemical transformations of other gas species in the air archive.\par

Indeed, some studies reported experimental evidence for a decreasing \RHe\ ratio in the atmosphere during the past few decades \citep[][see \secn{HeReview} for an extended literature review]{Sano:1988,Sano:1989,Sano:1998b}. While this decrease in the \RHe\ ratio was interpreted in terms of an anthropogenic input of fossil-fuel derived He into the atmosphere \citep{Sano:1988,Sano:1989,Sano:1998b}, other studies questioned the existence of a change in the atmospheric \RHe\ ratio \citep{Lupton:1991,Hoffmann:1993,Lupton:2004}. In light of the resulting controversy on the He isotope composition of the atmosphere \citep{Lupton:1991,Hoffmann:1993,Sano:1991}, recent studies attempted to further constrain the change in the \RHe\ ratio on longer time scales by analysing the \RHe\ ratio in air inclusions in ancient porcelain samples or metallurgical slags \citep{Pierson-Wickmann:2001,Matsuda:2010,Sano:2010}.\par

However, all experimental He isotope studies conducted so far used heterogeneous sets of air samples taken at different locations. An effect of geographical differences in the atmospheric \RHe\ ratio could therefore not be ruled out \citep{Sano:1988,Sano:1989,Sano:2010}. In addition, the \RHe\ data obtained from the air inclusions in ancient porcelain and slags may be affected by the release of radiogenic He isotopes from the porcelain or the slags into the air inclusions \citep{Pierson-Wickmann:2001,Matsuda:2010,Sano:2010}. Finally, the decrease rates of the \RHe\ ratio determined from the porcelain and slag samples depend strongly on poorly constrained assumptions on the timing and temporal evolution of the \RHe\ ratio in the atmosphere.\par

In summary, the knowledge on the elemental and isotopic noble gas composition of the atmosphere in the past is incomplete and the available He isotope data are inconsistent (see also \secn{HeReview}). While the concentrations of Ne, Ar, Kr and Xe isotopes in the atmosphere are generally assumed to be stable, there are no time series of measured data available that would allow a systematic assessment of this assumption. The He isotope data reported so far resulted in controversial findings on the potential change in the atmospheric He isotope composition in the past.\par

Robust and precise time series of atmospheric noble gas concentrations and isotope ratios would not only allow constraining  the noble gas isotope evolution in the atmosphere per se. New, robust He isotope data would also be useful in applications of the \RHe\ ratio as a proxy for environmental processes. For instance, the atmospheric \RHe\ ratio was discussed as a potential tracer to directly quantify the contribution of the mining and burning of fossil fuels to the accumulation of \methane, \COtwo\ and possibly also other gases in the atmosphere \citep{Sano:1993,Sano:2010}. The atmospheric \RHe\ ratio might even allow studying the effect of enhanced groundwater ventilation, which might be linked to global warming \citep{Pierson-Wickmann:2001}. Finally, better knowledge about the stability of the atmospheric noble gas composition would be highly useful to noble gas laboratories, because air is widely used as a standard gas for calibration of most noble gas analysis methods.\par

To determine the noble gas evolution of the atmosphere during the past few decades, we analysed the \NeTwenty, \ArFourty, \KrESix\ and \XeHTSix\ concentrations and the \RHe, \RHeNe, \RNe\ and \RAr\ ratios in the Cape Grim Air Archive (CGAA) \citep{Langenfelds:1996}. The CGAA was established with the specific aim of preserving a record of atmospheric composition, and has been used extensively for the reconstruction of a wide range of atmospheric trace gas histories \citep[e.g.,][]{ODoherty:2009,Muehle:2009,Muehle:2010,Vollmer:2011}. In contrast to the air archives used in previous He isotope studies, the CGAA tanks were always filled at the same location using consistent experimental methods. To determine a precise and robust record of the noble gas composition of the Cape Grim air during 1978--2011, we analysed the noble gas isotope composition of 80 air aliquots from replicate subsamples of six CGAA tanks.\par

\section{Potential changes of the atmospheric He composition during the past few decades}\seclabel{HeReview}
The mining and burning of fossil fuels has been postulated as the main source for an increasing He concentration in the atmosphere during the past few decades. The results of previous studies on the possible change in the atmospheric \RHe\ ratio are summarised in \tabl{RHe_rates}.\par

Modelling studies have predicted an increase in He concentration by 1--6\permille\ between 1939 and 1981 \citep{Oliver:1984}, and a rate of decrease in \RHe\ ratio during the past few decades of (0.14$\pm$0.07)--(0.85$\pm$0.3)\permille\ per year \citep[E\dwn{3}, F and G\dwn{1} in \tabl{RHe_rates};][]{Sano:1998b,Pierson-Wickmann:2001,Lupton:2004}.\par

Other potential He sources are the release of \HeThree\ from radioactive decay of synthetic \tritium\ used in nuclear weapons \citep{Lupton:2004}, and enhanced ventilation of terrigenic He from groundwaters due to the retreat of ice sheets and thawing of permafrost caused by global warming \citep{Pierson-Wickmann:2001}. These two processes are not considered further here because their predicted rates of change in the atmospheric \RHe\ ratio are at least two orders of magnitude lower than those predicted from the mining and burning of fossil fuels.\par

Trend analyses of \RHe\ ratios measured in archived air or in the deep water of the South Pacific indicated rates of decrease in the atmospheric \RHe\ ratio in the range of (0.79$\pm$0.6)--(2.14$\pm$0.6)\permille\ per year \citep[A, B and E\dwn{1,2} in \tabl{RHe_rates};][]{Sano:1988,Sano:1989,Sano:1998b}. However, results of other experimental analyses of \RHe\ ratios in air samples were interpreted to be consistent with a constant \RHe\ ratio in the atmosphere (C, D\dwn{1,2}, G\dwn{2,3} in \tabl{RHe_rates}), and indicated the decrease of the atmospheric \RHe\ ratio during 1973--2003 to be less than 0.1\permille\ per year on the 95\%  confidence level \citep{Lupton:1991,Lupton:2004}. This constraint is consistent with the decrease rate of  the atmospheric \RHe\ ratio predicted by a recent mass-balance model (F in \tabl{RHe_rates}), but not with the higher rates determined from other models (E\dwn{3} and G\dwn{1}) or measured data (A, B and E\dwn{1,2}).\par

In recent studies (H and I\dwn{1,2} in \tabl{RHe_rates}), \RHe\ ratios measured in air inclusions in porcelain samples dating back to 1400\,AD and in metallurgical slags dating back to 900\,AD were used to study the decrease rate of the atmospheric \RHe\ ratio on centennial time scales. However, the quantification of the change in the atmospheric \RHe\ ratio might be affected by the release of radiogenic He isotopes from the porcelain or slag matrices into the air inclusions. Furthermore, the rate of decrease in the \RHe\ ratio calculated from the \RHe\ ratios measured in the porcelain samples (H) is based on the arbitrary assumption of a constant \RHe\ ratio until 1750\,AD, followed by a linear increase of the \RHe\ ratio. However, this assumption is inconsistent with the \RHe\ ratios determined in the slag samples (I\dwn{1,2}), which suggest that the \RHe\ ratio has remained approximately constant until about 1900\,AD.\par

Up until now, all experimental determinations of the potential change in the atmospheric \RHe\ ratio during the past few decades used samples reflecting the atmospheric \RHe\ ratio at different geographical locations (\tabl{RHe_rates}). However, the atmospheric \RHe\ ratio may be variable on regional or global scales. For instance, a slightly higher \RHe\ ratio was observed in the southern hemisphere than in the northern hemisphere \citep{Sano:2010}, where most of the fossil fuels are mined and burned. Such geographical variations can therefore not be ruled out and may therefore present a possible source of error in the experimentally determined decrease rates of the atmospheric \RHe\ ratio during the past few decades \citep{Sano:1988,Sano:1989}.\par

\section{Methods}
\subsection{Cape Grim Air Archive (CGAA)}\seclabel{CGAA_archive}
The CGAA is a collection of gas tanks filled with air at the the Cape Grim Baseline Air Pollution Station, northwest Tasmania \citep{Langenfelds:1996}. Since 1978, more than 100 gas tanks have been filled with approximately 1--2\,\cubmSTP\ of pressurised air at intervals of about three months (22.414\,\cubmSTP $\equiv$ \E{3}\,mol). The tanks are filled under wind conditions where the trajectories of sampled air masses extend back over the `clean air sector' (190\deg--280\deg) of the Southern Ocean. This avoids contributions of mainland air masses that may be influenced by industrial or terrestrial exchange processes.\par

Cape Grim air is archived mostly in 35-L stainless steel tanks (Essex Cryogenics, Missouri, USA) which are internally electropolished and sealed with bellows-type valves. Since construction of the station building in the early 1980s, air has been drawn from the top of a 10\,m high tower through a stainless-steel  sampling tube (10\,cm outer diameter, OD) at a flow rate of approximately 0.3\,\cubmSTP/min. A previously flushed and evacuated tank is connected via thin tubing (1/4" OD) to a branch connection at the sampling tube. The tank is then cooled to 77\,K using liquid nitrogen, and the tank valve is opened to allow a split of the air flow through the sampling tube to expand into the tank. \Ntwo, \Otwo, \water\ and other gases are trapped in the tank by condensation on the cold tank surfaces, which maintains a steady air flow into the tank. The high gas-flow velocity in the thin tubing connecting the tank to the sampling tube prevents back flow of non-condensable gases (e.g., as verified by close agreement of \Htwo\ concentrations measured in CGAA tanks with those observed in CSIRO's parallel flask sampling program at Cape Grim). Filling of the archive tank is stopped after approximately 140\,min by closing the tank valve. After equilibration to ambient temperature, the tank is inverted such that the valve points downwards, which is then briefly opened to expel the liquid water out of the tank. The oldest (1978) sample was collected before the station building and 10\,m tower existed. This air was sampled from the cliff-top using a portable mast, with a gas flow of approximately 0.02\,\cubmSTP/min. The tank was flushed for 30\,min and then filled cryogenically using the same procedure as described above for later sampling events.\par

Assessment of the integrity of the archived air in the CGAA tanks is aided by past analyses of various trace gases and isotopes, such as the concentrations of \Htwo, \methane, \COtwo, \NtwoO\ and various halogenated compounds, \Otwo/\Ntwo\ and Ar/\Ntwo\ ratios, \isotope{N}{15}/\isotope{N}{14} ratio in \Ntwo\ and \isotope{O}{18}/\isotope{O}{16} ratio in \Otwo\ \citep{Langenfelds:1996,Langenfelds:2002,Langenfelds:2005,ODoherty:2009,Muehle:2009,Muehle:2010,Vollmer:2011}. In some cases these data have been validated by comparison to results from parallel, independent sampling programs \citep{Langenfelds:1996}. Some species have been measured repeatedly over many years providing insight into the stability of the archived air composition \citep{Langenfelds:1996}. This information is useful for identifying processes responsible for modification of the archived air in the tanks.\par

The CGAA tanks used for this study were selected to (i) be evenly distributed over the longest possible period (1978--2011) and (ii) have exhibited maximum integrity of trace gas composition with respect to processes that could influence the noble gas composition of the archived air. Because the noble gases are chemically inert, the main aim was to avoid artefacts resulting from mass-dependent fractionation. Previous studies showed that some of the oldest batch of 35-L CGAA tanks used until 1988 had developed microscopic leaks at welded sites on their surfaces, resulting in small leaks and varying degrees of mass-dependent fractionation of the stored air \citep{Langenfelds:2002}. A direct indicator of such fractionation available to us was from CGAA measurements of \isotope{N}{15}/\isotope{N}{14} ratios in \Ntwo\ and related species made in 2001 \citep{Langenfelds:2002}. Based on these data we selected a subset of CGAA tanks containing air that was expected to be negligibly fractionated for the purposes of the present study. All but one of the selected CGAA tanks used in this study showed no detectable anomaly in \isotope{N}{15}/\isotope{N}{14} ratios in \Ntwo\ with values within $\pm0.005$\permille\ of modern atmospheric air. Only one tank (filled in 1984) showed a marginally significant increase of the \isotope{N}{15}/\isotope{N}{14} ratio by 0.012\permille\ in 2001. While CSIRO is not equipped to repeat the \isotope{N}{15}/\isotope{N}{14} analyses at this time, subsequent trace-gas analyses of the air in this tank at CSIRO suggest that between 2001 and 2011 the rate of fractionation has increased. Significant variations are evident in the concentrations of \Htwo\ (22\permille\ decrease during 2001--2011), \methane\ (2.1\permille\ decrease), \COtwo\ (1.4\permille\ increase) and \NtwoO\ (2.0\permille\ increase). The relative concentration variations among these species are consistent to within measurement uncertainty with fractionation due to gas molecules escaping through a narrow orifice whose diameter is small compared to the mean free path length of inter-molecular collisions (effusion). Graham's Law states that the rate of effusion of a given gas species from the tank is inversely proportional to the square root of the molecular mass of this species. We will therefore use GrahamÕs Law to assess possible noble gas fractionation in the 1984 tank. Recent CSIRO analyses of \Htwo, \methane, \COtwo\ and \NtwoO\ show no evidence of any significant fractionation in the other five CGAA tanks considered here.\par

\subsection{CGAA subsampling and noble gas analysis}
Subsamples for noble gas analysis were transferred from six CGAA tanks (filled in 1978, 1984, 1993, 2004, 2010 and 2011; \tabl{NG_data}) into 1.2\,m long copper tubes (3/8" OD) at CSIRO Aspendale (Australia). The CGAA tanks were fitted with internally electropolished, stainless steel regulators (Tescom), separated from the tank valve by a 0.6\,m length of 1/16" OD stainless steel tubing. A 1\,m length of 1/16" OD stainless steel tubing connected the regulator to the copper tube. A further 1\,m length of 1/16" OD stainless steel tubing was attached to the outlet of the copper tube to prevent back diffusion of ambient air into the copper tube during subsampling, and led to a manifold fitted with a pressure gauge, needle valve and flowmeter to control gas flow during sampling. Before gas transfer from the tank into the copper tube, the copper tube was evacuated to remove ambient air using an electric diaphragm pump attached to the manifold. Air flow was then commenced, and maintained at a rate of about 20\,\ccSTP/min with air pressure in the copper tube held steady at 145\,kPa for a flushing period of 10--30\,min. These flow parameters were selected to minimise the potential for mass-dependent fractionation, as guided by results of gas transfer tests reported previously \citep{Langenfelds:2005}. Finally the copper tube was sealed at both ends using a crimping tool. Multiple replicate subsamples were taken from each CGAA tank.\par

The noble gas isotope compositions in the CGAA subsamples (copper tubes) were analysed in the Noble Gas Laboratory at ETH Zurich (Switzerland). Multiple air aliquots (0.04--0.7\,\ccSTP) from each subsample were analysed by static mass spectrometry \citep{Beyerle:2000}. After transferring an aliquot into the gas purification system, the copper tube was re-sealed, and the air aliquot was separated into a He-Ne fraction and an Ar-Kr-Xe fraction in the purification system using a series of cold traps cooled by liquid nitrogen. A first split of the He-Ne fraction was used for combined analysis of \HeFour, \NeTwenty, and \NeTwentyTwo\ using a custom-built sector-field mass spectrometer equipped with a Baur-Signer ion source \citep{Baur:1980}. After He-Ne analysis, the Ar-Kr-Xe fraction was used for combined analysis of \ArTSix, \ArFourty, \KrESix, and \XeHTSix\ using the same mass spectrometer. A second split of the He-Ne fraction was further purified using a cryo trap at 70\,K and subsequently used for \RHe\ analysis in a Micromass 5400 sector-field mass spectrometer equipped with a Nier-type ion source. The isotope concentrations and ratios measured in the different gas aliquots from the same subsample were independent of the aliquot size, because the design of the Baur-Signer ion source inherently minimises changes in sensitivity as a function of the gas pressure in the mass spectrometer, and the Nier-type ion source is tuned for gas-pressure independent \RHe\ ratio measurement instead of maximum sensitivity. Correction of non-linearities of the mass-spectrometer results obtained from different gas amounts (\tabl{NG_data}) was therefore not necessary. Mass spectrometric results were calibrated using aliquots of an air standard collected in Zurich (Switzerland) on 7 June 1996 (\tabl{slowcals}), which were processed exactly in the same way as the CGAA subsample aliquots.\par

The standard errors for the noble gas concentrations and isotope ratios determined in the individual gas aliquots correspond to the standard deviations of the results of the standard analyses \citep{Beyerle:2000}, which are listed in \tabl{slowcals}. We found that the total analytical precision is controlled mainly by the preparation and cleaning of the gas aliquot, whereas the uncertainties of the gas analysis in the mass spectrometers are of minor importance on the overall precision. To improve precision and robustness of the results of each CGAA tank, several gas aliquots were analysed from each subsample. To take into account potential effects associated with the gas transfer from the CGAA tanks into the subsample tubes on the overall data precision, air aliquots were taken from different subsample replicates of the same CGAA tank. Data values corresponding to the same CGAA tank were then combined by calculating their means and errors of the means.\par

\section{Results and Discussion}
For each CGAA tank, at least nine air aliquots were analysed. These air aliquots were taken from at least two different subsample replicates, respectively. Only for the 2010 tank, all aliquots were taken from the same subsample. The noble gas results from each subsample aliquot as well as the means and standard errors of the means of each CGAA tank are listed in \tabl{NG_data}.\par

The subsamples from the 2010 tank and some of the subsamples of the 2004 tank were filled in a laboratory where He is released into the ambient air from gas chromatographic instruments. An ambient air sample collected in this laboratory showed a He concentration of 35\,\ppmV, which is about 7 times higher than in uncontaminated air. The concentrations of the other noble gases in the laboratory air agreed with those in uncontaminated air. The \RHeNe\ ratio determined in the subsamples filled in this laboratory was about 1.25\%  higher than in replicates taken in a room with uncontaminated ambient air. 
Therefore, if the subsamples are assumed to reflect a binary mixture of air from the CGAA tanks and a small amount of He-contaminated ambient air, contamination with ambient air during gas transfer from the tanks into the copper tubes can be constrained to $\lesssim$0.2\%\ for all subsamples. To avoid a bias of the He isotope composition determined in the 2004 and 2010 CGAA tanks, the He data from the subsamples filled in the gas-chromatography laboratory were not used (`NA' values in \tabl{NG_data}).\par

\subsection{Stability of neon, argon, krypton and xenon}
\figr{NeArKrXe} shows the time series of the concentrations of \NeTwenty, \ArFourty, \KrESix\ and \XeHTSix\ and the isotope ratios \RNe\ and \RAr\ in the CGAA. The potential effect of effusion due to an assumed leak in the 1984 tank as calculated from Graham's Law is similar to the analytical precision (\figr{NeArKrXe}, dashed extensions of the error bars). The additional uncertainty associated with the potential gas fractionation by effusion from this tank is therefore only marginal.\par

The \NeTwenty, \ArFourty, \KrESix, \XeHTSix, \RNe\ and \RAr\ time series do not show any systematic changes or trends (see also statistical test results in \figr{NeArKrXe}). This finding agrees with the assumption that the Ne, Ar, Kr and Xe composition of the Cape Grim air was stable during the past few decades. Also, the error-weighted means of the isotope concentrations and ratios in the CGAA ($M$ in \figr{NeArKrXe}) agree with the composition of the Zurich air used for calibration of the data (\tabl{slowcals}). While these findings were to be expected based on theoretical considerations, they are the first systematic and experimental validation of the temporal and spatial uniformness of the \NeTwenty, \NeTwentyTwo, \ArTSix, \ArFourty, \KrESix\ and \XeHTSix\ composition of the atmosphere during the past few decades.\par

\subsection{Variability of helium isotopes}
\figr{He} shows the time series of the \RHe\ ratio in the CGAA. The potential effusion effect on the \RHe\ ratio in the 1984 tank is similar to the analytical precision. A $\chi^2$ test of the `no trend' hypothesis (line A in \figr{He}) yields $p = 0.05$ using the uncorrected data or $p = 0.005$ using the effusion-corrected data. Without considering alternative hypotheses, the `no trend' hypothesis can therefore only be rejected on the 2$\sigma$ level using the effusion-corrected data. However, fitting a straight line to the \RHe\ ratio time series by error-weighted least-squares regression (line B in \figr{He}) results in a slope that corresponds to a mean rate of decrease in the \RHe\ ratio of $(0.23\pm0.08)$\permille\ per year for the uncorrected data or $(0.30\pm0.08)$\permille\ per year for the effusion-corrected data (1$\sigma$ errors). These two values agree to within their standard errors, and both values are not consistent with a constant \RHe\ ratio during 1978--2011 on the 2$\sigma$ level. In addition, a statistical comparison of the `linear trend' and `no trend' models \citep[$F$-test using the error-weighted sum of squared residuals;][]{Faraway:2005} shows clearly that the `no trend' model should be rejected in favour of the `linear trend' model ($p = 0.0031$ for the uncorrected data and $p = 0.038$ for the effusion-corrected data). Our He isotope results therefore provide strong evidence for a slight decrease in the atmospheric \RHe\ ratio during the past few decades.\par

The rate of decrease in the \RHe\ ratio determined from our noble gas isotope record is consistent with all previously determined rates within their 2$\sigma$ error limits, except rates A, G\dwn{2,3} and I\dwn{1,2} (\tabl{RHe_rates}, \figr{RHe_comparison}). Rate A was determined from \RHe\ ratios measured in different samples taken at different locations and analysed in different laboratories using different instruments and experimental protocols. The rates G\dwn{2} and G\dwn{3} were determined from \RHe\ ratios measured in 5 (G\dwn{2}) or 7 (G\dwn{3}) air samples, which were taken at different geographical locations. These rates also seemed to depend on the standards used for data calibration. The \RHe\ ratio decrease rates A, G\dwn{2} and G\dwn{3} may therefore be affected by differences in the samples, methods and instruments used. The \RHe\ ratio decrease rates I\dwn{1} and I\dwn{2} were determined from \RHe\ ratios measured in ancient slags. These \RHe\ ratios may be affected by the release of radiogenic He isotopes from the slag matrix into the air inclusions. In addition, these data represent the atmospheric \RHe\ ratio before or during the early stages of the industrial revolution and are therefore expected to yield a lower rate of decrease in the \RHe\ ratio than the rate reflecting the past few decades, when most of the fossil fuels were mined and burned. In contrast, the rate of decrease in the \RHe\ ratio determined in our study is unaffected by geographical differences of the \RHe\ ratio and is based on robust \RHe\ ratios determined in a reliable air archive covering the time range of 1978--2011.\par

The \RHeNe\ ratios we determined in the CGAA are much more precise than  than \RHeNe\ ratios determined in previous air archive studies. Our \RHeNe\ data (\figr{HeNe}) might therefore be useful to further constrain the atmospheric \HeFour\ evolution during the past few decades. For instance, if the decrease in the \RHe\ ratio is indeed due to an increased addition of He with a low \RHe\ ratio to the atmosphere (e.g., as a result of the mining and burning of fossil fuels), the \RHeNe\ ratio would be expected to increase at a similar rate as the \RHe\ ratio decreases (dashed trend line in \figr{HeNe}). This hypothesis cannot be ruled out based on the observed \RHeNe\ ratios. However, due to the large ratios of the masses of \HeFour\ and \NeTwenty, the potential effect of effusion on the \RHeNe\ ratio in the 1984 tank is large. Given this considerable uncertainty of the \RHeNe\ ratio in the 1984 tank, we refrain from deriving further constraints of the atmospheric \HeFour\ evolution using our \RHeNe\ data.\par

\section{Conclusions}
Our noble gas analyses of 80 air aliquots taken from replicate subsamples of six CGAA tanks yielded robust and precise \NeTwenty, \ArFourty, \KrESix\ and \XeHTSix\ concentrations and \RHe, \RNe\ and \RAr\ ratios in Cape Grim air during 1978--2011. The noble gas composition of the originally sampled air is well preserved in the CGAA. In contrast to the archives used in previous studies, potential artefacts in the noble gas isotope composition resulting from sampling or storage are well constrained from independent trace-gas monitoring data, and might be relevant only in case of the 1984 tank. The  noble gas record determined in the CGAA therefore allowed us to determine the first systematic experimental evidence supporting the general assumption of stable Ne, Ar, Kr and Xe concentrations in the atmosphere during the past few decades.\par

Our He isotope results provide strong evidence for a decrease of the atmospheric \RHe\ ratio at Cape Grim during 1978--2011. The mean rate of decrease amounts to $(0.23\pm0.08)$\permille\ per year, or to $(0.30\pm0.08)$\permille\ per year if a potential effusion effect due to an assumed leak in the 1984 tank is included in the analysis (1$\sigma$ errors). These rates are consistent with many, but not all, previous determinations of the decrease of the atmospheric \RHe\ ratio during the past few decades. However, in contrast to previously measured decrease rates, our result is based on a large and robust data set of \RHe\ ratios measured in high-quality samples of archived air which was always taken at the same location and under the same wind conditions. Geographical variabilities in the atmospheric \RHe\ ratio can therefore be ruled out as a possible source of error in our analysis of the change in the atmospheric \RHe\ ratio. The decrease rate of the atmospheric \RHe\ ratio determined here therefore provides new and more robust constraints on the evolution of the atmospheric \RHe\ ratio during the past few decades. While it was beyond the scope of our study to fully elucidate the sources and processes resulting in the observed change of the \RHe\ ratio, our result is consistent with most predictions of the change in the atmospheric \RHe\ ratio as a result from the mining and burning of fossil fuels.\par

To further investigate the nature and the source of the changing He isotope composition in the atmosphere, it might be useful to further improve the analytical precision of the \RHe\ ratio \citep[e.g.,][]{Sano:2008,Mabry:2010,Jean-Baptiste:2012}, as well as the temporal resolution of the time series in the \RHe\ and \RHeNe\ ratios by analysing subsamples from additional CGAA tanks. Combining precise and high-resolution time series of \RHe\ and \RHeNe\ ratios are expected to yield further insights and quantitative constraints with respect to the timing and the dynamics of the (possibly anthropogenic) sources and processes resulting in the variability of the atmospheric He isotope composition.\par

 \section*{Acknowledgements}
The Cape Grim Air Archive Program is supported by the Australian Bureau of Meteorology. We thank the Cape Grim staff who over many years have been involved in the collection and management of this air archive.
 
\clearpage
\singlespacing

\bibliographystyle{elsarticle-harv} 
\bibliography{MB_Literatur}

\nolinenumbers

\clearpage

\section*{Figures}
\pagestyle{empty}

\begin{figure}[p]
\begin{center}
$\begin{array}{cc}
\includegraphics[width=\textwidth]{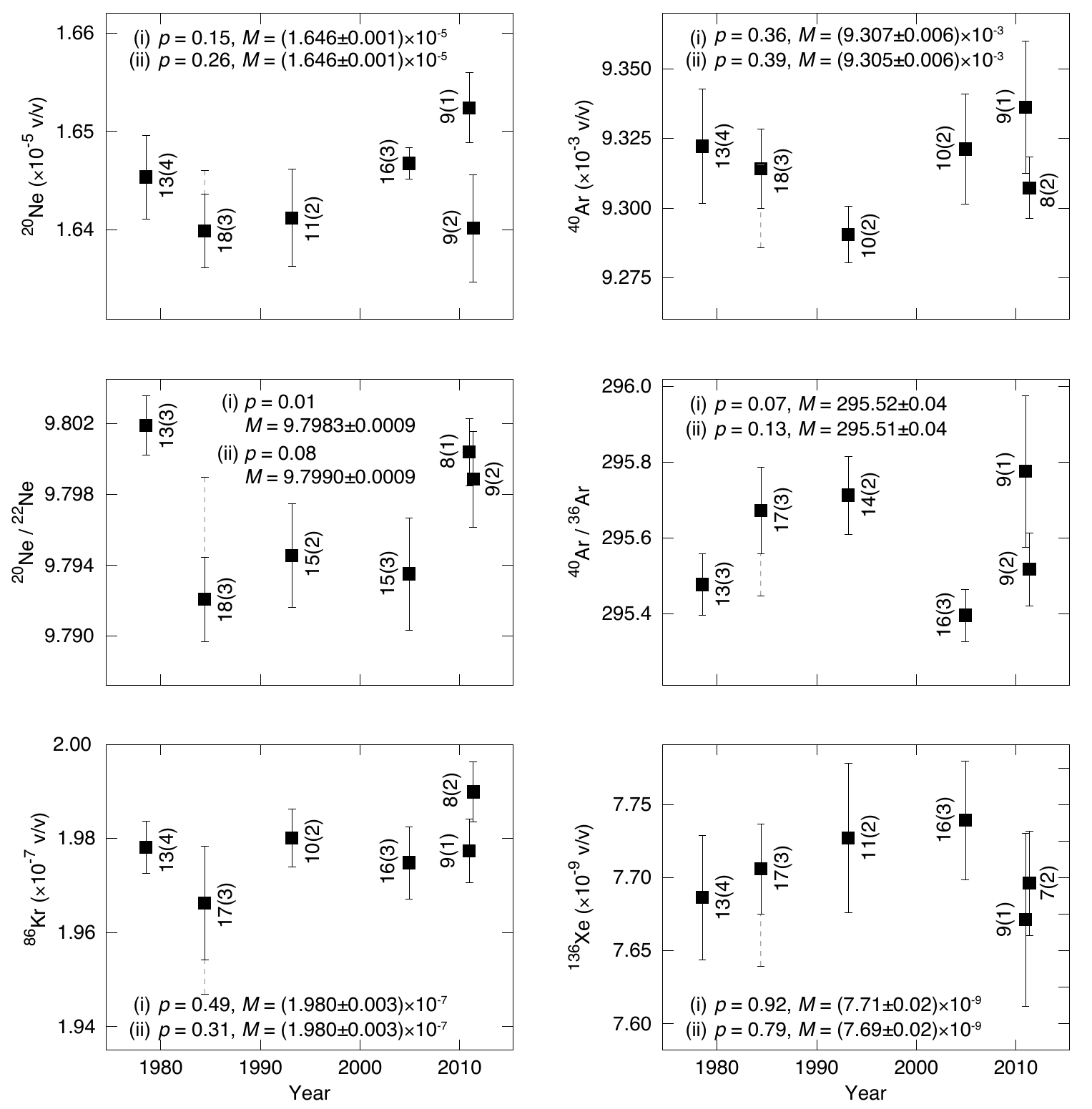}
\end{array}$
\end{center}
\caption{Volumetric concentrations of \NeTwenty, \ArFourty, \KrESix\ and \XeHTSix\ and isotope ratios \RNe\ and \RAr\ in Cape Grim Air Archive tanks (means and standard errors of the means as given in \tabl{NG_data}). The dashed extensions of the error bars in the 1984 data indicate the potential effusion effect in the 1984 tank (see text). $p$ is the $p$-value resulting from a $\chi^2$ test that the values plotted in each panel can be described by a common mean value, and $M$ is the error-weighted mean of the data plotted in each panel. $p$ and $M$ were both calculated by (i) excluding or (ii) including the potential effusion effect in the 1984 data. Numbers next to the squares indicate the number of aliquots and subsamples (in parentheses) used to determine each data value.}
\figlabel{NeArKrXe}
\end{figure}

\clearpage

\begin{figure}[p]
\begin{center}
$\begin{array}{cc}
\includegraphics[width=0.7\textwidth]{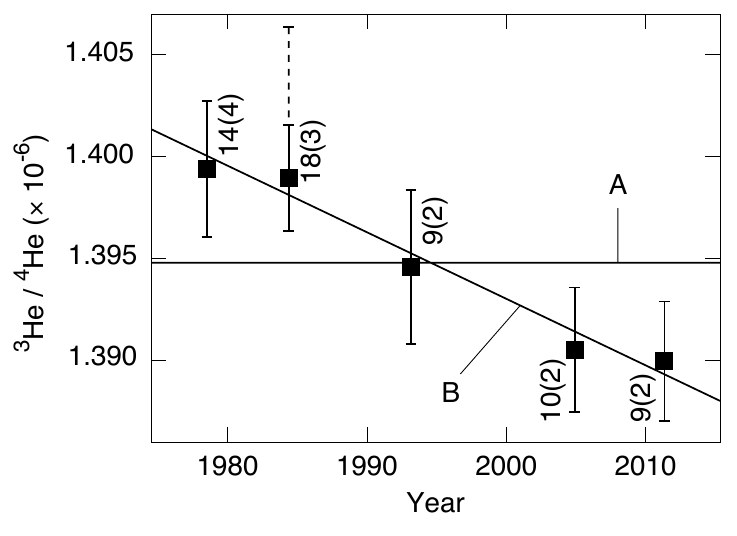}
\end{array}$
\end{center}
\caption{\RHe\ ratio observed in Cape Grim Air Archive tanks (means and standard errors of the means as given in \tabl{NG_data}). The dashed extension of the error bar in the 1984 value indicates the potential effusion effect in the 1984 tank (see text). Line A reflects the error-weighted mean value (`no trend' model excluding the potential effusion effect). Line B reflects least squares regression of a straight line (`linear trend' model excluding the potential effusion effect). Numbers next to the squares indicate the number of aliquots and subsamples (in parentheses) used to determine each data value.}
\figlabel{He}
\end{figure}

\clearpage

\begin{figure}[p]
\begin{center}
$\begin{array}{cc}
\includegraphics[width=0.7\textwidth]{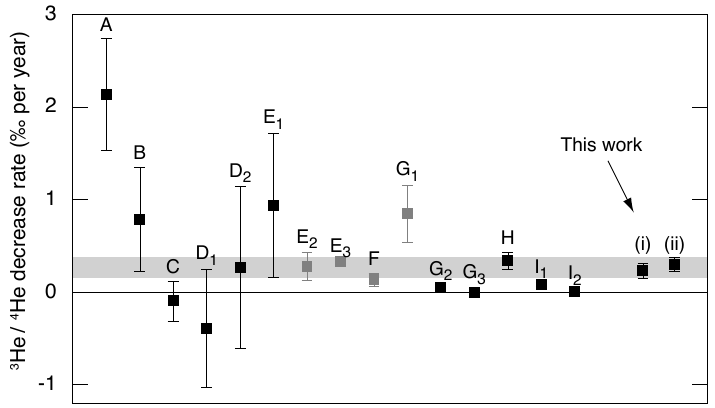}
\end{array}$
\end{center}
\caption{Comparison of rates of decrease in the atmospheric \RHe\ ratio reported in the literature (\tabl{RHe_rates}) and the rates determined in this work by (i) excluding or (ii) including the effusion correction of the data from 1984 tank in the regression of the trend line. Results are shown in chronological sequence of their publication. Black squares correspond to measured values, grey squares to modelled values. Error bars reflect standard errors (see notes in \tabl{RHe_rates}). The horizontal grey bar indicates the range of the rates determined in this work including their standard errors.}
\figlabel{RHe_comparison}
\end{figure}

\clearpage

\begin{figure}[p]
\begin{center}
\includegraphics[width=0.7\textwidth]{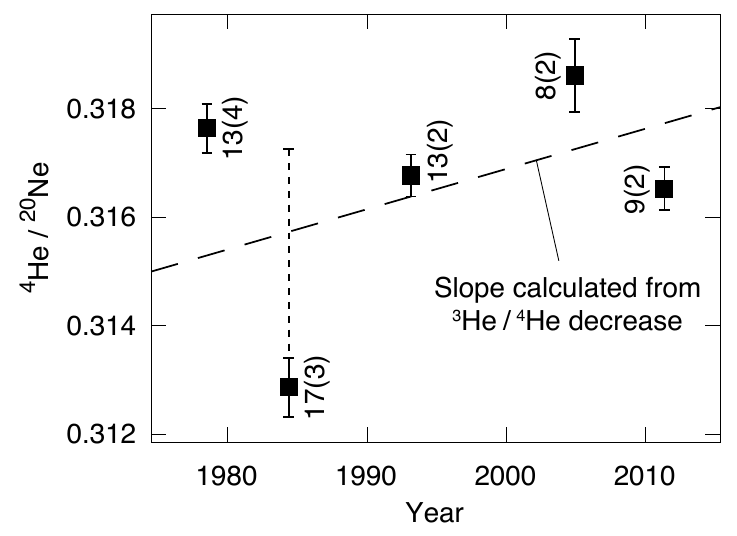}
\end{center}
\caption{\RHeNe\ ratio in Cape Grim Air Archive tanks (means and standard errors of the means as given in \tabl{NG_data}). The dashed extension of the error bar in the 1984 value indicates the potential effusion effect in the 1984 tank (see text). The dashed trend line indicates the linear trend resulting from the assumption that the rate of increase in the \RHeNe\ ratio equals that of the decrease in the \RHe\ ratio (calculated by excluding the potential effusion effect in the 1984 data; see text). Numbers next to the squares indicate the number of aliquots and subsamples (in parentheses) used to determine each data value.}
\figlabel{HeNe}
\end{figure}

\clearpage

\section*{Tables}
\begin{table}[p]
\caption{Summary of previously reported rates of decrease in the \RHe\ ratio ($r$) and their standard errors, listed in chronological sequence of their publication.}
\setlength{\tabcolsep}{1pt}
\footnotesize
\begin{tabular}{c c p{0.67\linewidth} p{0.2\linewidth}}
\toprule
Label & $r$ & Descripton & Ref.\\
&  (\permille\ per year) \\
\midrule
A & 2.14$\pm$0.6 & Comparison of absolute determinations of the \RHe\ ratios in three different air samples taken at Leningrad (1969), Ontario (1975 or earlier) and Tokyo (1988), each analysed in a different laboratory (we assumed the reported error to reflect the 1$\sigma$ level) & \cite{Sano:1988}\\
B & 0.79$\pm$0.6 &  Trend analysis of \RHe\ ratios measured in 20 air samples taken at different locations in Japan and in North America (California, USA) during 1977--1988 (we assumed the reported error to reflect the 1$\sigma$ level) &  \cite{Sano:1989}\\
C & -0.09$\pm$0.2 & Trend analysis of \RHe\ ratios measured in 5 air samples taken at different locations at the east coast of North America (California, USA) during 1973--1990 & \cite{Lupton:1991}\\
D\dwn{1} & -0.39$\pm$0.64 &  Comparison of absolute determinations of the \RHe\ ratios in four different air samples taken at Minneapolis (USA, 1956), Leningrad (1969), Ontario (1975 or earlier) and Tokyo (1988), each analysed in a different laboratory (we assumed the reported error to reflect the 1$\sigma$ level) & \cite{Hoffmann:1993}\\
D\dwn{2} & 0.27$\pm$0.87 & Same as D\dwn{1}, but excluding the value determined from the sample taken in Tokyo (1988) & \cite{Hoffmann:1993}\\
E\dwn{1} & 0.94$\pm$0.78 & Based on \RHe\ ratios measured in deep-water samples from the South Pacific & \cite{Sano:1998b}\\
E\dwn{2} & 0.28$\pm$0.15 & Calculated by combining previously measured values & \cite{Sano:1998b}\\
E\dwn{3} & 0.33 & Modeled value based on estimated He release from mining and burning of fossil fuels (error was not estimated) & \cite{Sano:1998b}\\
F & 0.14$\pm$0.07 & Modeled value based on estimated He release from mining and burning of fossil fuels (we assumed the reported error to reflect the 1$\sigma$ level). In the same study, \RHe\ ratios in air inclusions in ancient metallurgical slags were analyses. However, no rate of decrease in the \RHe\ ratio could be determined from these results. & \cite{Pierson-Wickmann:2001}\\
G\dwn{1} & 0.85$\pm$0.3 & Calculated from predicted He concentration increase from 1939 to 1981 resulting from mining and burning of fossil fuels \citep{Oliver:1984} & \cite{Lupton:2004}\\
G\dwn{2} & 0.05$\pm$0.03 & Trend analysis of 7 \RHe\ ratios measured in 7 air samples taken at different locations at the east coast of North America (California and Oregon, USA) during 1973--2003 & \cite{Lupton:2004}\\
G\dwn{3} & 0.00$\pm$0.05 & Same as G\dwn{2}, but using only the data from 5 samples, which were evaluated using a different gas standard & \cite{Lupton:2004}\\
H   & 0.34$\pm$0.09 & Trend analysis of \RHe\ ratios measured in air inclusions in porcelain samples originating from different locations in Japan and China, dating back to about 1400\,AD, including a modern porcelain sample & \cite{Matsuda:2010}\\
I\dwn{1} & 0.08$\pm$0.02 & Trend analysis of \RHe\ ratios measured in air inclusions in metallurgical slags originating from different locations in Japan, dating back to about 900--1900\,AD, and a modern air sample & \cite{Sano:2010}\\
I\dwn{2} & 0.00$\pm$0.03 & Same as I\dwn{1}, but the value of the \RHe\ ratio of the modern air sample was excluded from the evaluation & \cite{Sano:2010}\\
\bottomrule
\end{tabular}
\tablabel{RHe_rates}
\end{table}
\clearpage

\begin{table}[p]
\caption{Standard gas aliquots: relative standard deviations of the air standard results ($\sigma^*_i$) and assumed isotope composition of our air standard (see text; $R^*_i$: isotope ratios, $C^*_i$:  volumetric concentration).}
\tablabel{slowcals}
\begin{center}
\begin{tabular}{c c c c}
\toprule
$i$ & $\sigma^*_i$ & $R^*_i$ & $C^*_i$ (v/v) \\
\midrule
\RHe & 0.6\% & 1.399\Edot{-6}\\
\RHeNe & 0.6\% & 0.318\\
\RNe & 0.1\% & 9.80 \\
\RAr & 0.09\% & 295.5$^\dagger$ \\
\NeTwenty & 1.4\% & & 1.645\Edot{-5} \\
\ArFourty & 0.3\% & & 9.304\Edot{-3}\\
\KrESix & 1.6\% & & 1.984\Edot{-7} \\
\XeHTSix & 1.9\% & & 7.71\Edot{-9}\\
\bottomrule
\multicolumn{4}{p{9cm}}{\footnotesize $^\dagger$: For consistency with previous data measured in our noble gas laboratory, we use the \RAr\ ratio recommended by the IUGS Subcommission on Geochronology \citep{Steiger:1977} rather than the more recent value of 298.6 \citep{Lee:2006,Mark:2011}.}
\end{tabular}
\end{center}
\end{table}

\clearpage

{
\footnotesize
\setlength{\tabcolsep}{3pt}
\begin{longtable}{c c c c c c c c c c}
\caption{Noble gas isotope ratios and concentrations in individual subsample aliquots ($S$: subsample number, $A$: gas amount in aliquot, $\mu$: means of values corresponding to a given tank, $\sigma_\mu$: standard errors of means). Missing values (NA) are due to technical problems during analysis or potential subsample contamination by He (see text).}\tablabel{NG_data} \\ 
\toprule
$S$ & $A$ & \RHe & \RHeNe & \RNe & \RAr & \NeTwenty & \ArFourty & \KrESix & \XeHTSix \\ 
 & (\ccSTP) & ($\times$\E{-6}) & & & & ($\times$\E{-5}v/v) & ($\times$\E{-3}v/v) & ($\times$\E{-7}v/v) & ($\times$\E{-9}v/v) \\ 
\midrule
\endfirsthead
\caption{Noble gas isotope ratios and concentrations in individual subsample aliquots (continued).} \\ 
\toprule
$S$ & $A$ & \RHe & \RHeNe & \RNe & \RAr & \NeTwenty & \ArFourty & \KrESix & \XeHTSix \\ 
 & (\ccSTP) & ($\times$\E{-6}) & & & & ($\times$\E{-5}v/v) & ($\times$\E{-3}v/v) & ($\times$\E{-7}v/v) & ($\times$\E{-9}v/v) \\ 
\midrule
\endhead
\multicolumn{10}{c}{\emph{Cape Grim Air Archive tank: 7.7.1978 (UAN: 780002)}} \\ 
1 & 0.641 & 1.381 &     NA & 9.793 & 295.2 & 1.62 & 9.26 & 2.00 & 7.58 \\ 
1 & 0.323 & 1.409 & 0.3175 & 9.798 & 295.4 & 1.64 & 9.31 & 1.99 & 7.68 \\ 
1 & 0.155 & 1.416 & 0.3172 & 9.799 & 295.2 & 1.64 & 9.33 & 1.99 & 7.52 \\ 
1 & 0.627 & 1.394 & 0.3171 & 9.803 & 295.3 & 1.63 & 9.20 & 1.94 & 7.40 \\ 
1 & 0.609 & 1.400 & 0.3161 & 9.801 & 295.2 & 1.65 & 9.23 & 1.97 & 7.88 \\ 
2 & 0.548 & 1.394 & 0.3161 & 9.799 & 295.6 & 1.64 & 9.24 & 1.95 & 7.58 \\ 
2 & 0.601 & 1.408 & 0.3187 & 9.814 & 295.7 &   NA &   NA &   NA &   NA \\ 
2 & 0.265 & 1.419 & 0.3154 & 9.800 & 295.4 & 1.66 & 9.31 & 1.99 & 7.72 \\ 
2 & 0.116 & 1.395 & 0.3157 & 9.797 & 295.5 & 1.67 & 9.39 & 1.97 & 7.76 \\ 
2 & 0.047 & 1.379 & 0.3178 & 9.812 & 296.2 & 1.67 & 9.38 & 1.96 & 7.92 \\ 
3 & 0.118 & 1.391 & 0.3206 &    NA &   NA & 1.64 & 9.37 & 1.97 & 7.57 \\ 
4 & 0.567 & 1.389 & 0.3192 & 9.801 & 295.8 & 1.62 & 9.32 & 1.98 & 7.71 \\ 
4 & 0.550 & 1.409 & 0.3181 & 9.808 & 295.6 & 1.65 & 9.39 & 2.01 & 7.90 \\ 
4 & 0.526 & 1.407 & 0.3198 & 9.802 & 295.3 & 1.66 & 9.45 & 2.00 & 7.70 \\ 
\midrule
& \it $\mu$: & 1.399 & 0.3176 & 9.802 & 295.5 & 1.645 & 9.32 & 1.98 & 7.69 \\ 
& $\sigma_\mu$: & 0.003 & 0.0004 & 0.002 &  0.1 & 0.004 & 0.02 & 0.01 & 0.04 \\ 
\midrule 
\multicolumn{10}{c}{\emph{Cape Grim Air Archive tank: 23.5.1984 (UAN: 840004)}} \\ 
1 & 0.598 & 1.393 & 0.3136 & 9.792 & 295.3 & 1.63 & 9.30 & 1.99 & 7.81 \\ 
1 & 0.267 & 1.405 & 0.3131 & 9.800 & 295.8 & 1.64 & 9.36 & 2.01 & 7.69 \\ 
1 & 0.106 & 1.401 & 0.3142 & 9.780 & 296.7 & 1.65 & 9.40 & 2.01 & 7.66 \\ 
1 & 0.038 & 1.374 & 0.3173 & 9.795 &   NA & 1.63 & 9.29 & 1.98 & 7.82 \\ 
1 & 0.624 & 1.396 & 0.3122 & 9.798 & 295.7 & 1.68 & 9.38 & 2.01 & 7.73 \\ 
1 & 0.600 & 1.404 & 0.3125 & 9.797 & 295.3 & 1.64 & 9.30 & 1.99 & 7.68 \\ 
2 & 0.648 & 1.386 & 0.3140 & 9.796 & 295.5 & 1.63 & 9.27 & 1.97 & 7.59 \\ 
2 & 0.334 & 1.400 &     NA & 9.806 & 295.8 & 1.64 & 9.27 & 1.95 & 7.73 \\ 
2 & 0.160 & 1.407 & 0.3136 & 9.795 & 294.7 & 1.64 & 9.35 & 1.85 & 7.68 \\ 
2 & 0.071 & 1.399 & 0.3157 & 9.801 & 295.7 & 1.65 & 9.45 & 1.98 & 7.60 \\ 
2 & 0.624 & 1.384 & 0.3136 & 9.787 & 295.7 & 1.65 & 9.25 & 1.97 & 7.86 \\ 
2 & 0.313 & 1.394 & 0.3157 & 9.778 & 296.4 & 1.61 & 9.26 & 1.91 & 7.38 \\ 
3 & 0.557 & 1.401 & 0.3125 & 9.787 & 295.8 & 1.61 & 9.19 & 1.98 & 7.70 \\ 
3 & 0.534 & 1.403 & 0.3105 & 9.803 & 295.6 & 1.65 & 9.34 & 1.91 & 7.88 \\ 
3 & 0.234 & 1.421 & 0.3104 & 9.775 & 296.1 & 1.62 & 9.32 & 1.89 & 7.55 \\ 
3 & 0.226 & 1.417 & 0.3091 & 9.803 & 295.0 & 1.66 & 9.30 & 2.00 & 7.80 \\ 
3 & 0.430 & 1.400 & 0.3101 & 9.792 & 295.7 & 1.64 & 9.30 & 2.05 & 7.83 \\ 
3 & 0.401 & 1.397 & 0.3105 & 9.773 & 295.7 & 1.64 & 9.33 &   NA &   NA \\ 
\midrule
& \it $\mu$: & 1.399 & 0.3129 & 9.792 & 295.7 & 1.640 & 9.31 & 1.97 & 7.71 \\ 
& $\sigma_\mu$: & 0.003 & 0.0005 & 0.002 &  0.1 & 0.004 & 0.01 & 0.01 & 0.03 \\ 
\midrule 
\multicolumn{10}{c}{\emph{Cape Grim Air Archive tank: 2.3.1993 (UAN: 930279)}} \\ 
1 & 0.530 & 1.388 & 0.3191 & 9.814 & 295.5 & 1.68 & 9.35 &   NA & 7.62 \\ 
1 & 0.223 &    NA & 0.3161 & 9.804 & 295.9 &   NA &   NA &   NA &   NA \\ 
1 & 0.244 &    NA &     NA & 9.792 & 295.5 &   NA &   NA &   NA &   NA \\ 
1 & 0.106 &    NA & 0.3163 & 9.781 & 296.6 &   NA &   NA &   NA &   NA \\ 
1 & 0.116 &    NA &     NA & 9.788 & 295.1 &   NA &   NA &   NA &   NA \\ 
1 & 0.603 & 1.391 & 0.3166 & 9.803 & 295.6 & 1.63 &   NA & 1.97 & 7.75 \\ 
1 & 0.293 & 1.405 & 0.3166 & 9.800 & 295.3 & 1.65 & 9.27 & 1.95 & 7.67 \\ 
1 & 0.165 & 1.410 & 0.3157 & 9.797 & 296.1 & 1.63 & 9.25 & 1.96 & 7.97 \\ 
1 & 0.088 & 1.409 & 0.3143 & 9.779 & 296.2 & 1.63 & 9.26 & 1.96 & 7.64 \\ 
1 & 0.060 & 1.386 & 0.3183 & 9.783 &   NA & 1.66 & 9.33 & 1.99 & 7.38 \\ 
2 & 0.673 & 1.390 & 0.3156 & 9.797 & 295.6 & 1.64 & 9.28 & 2.00 & 7.76 \\ 
2 & 0.654 & 1.377 & 0.3176 & 9.815 & 295.5 & 1.63 & 9.28 & 2.00 & 7.78 \\ 
2 & 0.627 & 1.395 & 0.3174 & 9.783 & 295.5 & 1.63 & 9.29 & 2.01 & 8.02 \\ 
2 & 0.571 &    NA & 0.3188 & 9.786 & 295.8 & 1.63 & 9.28 & 1.98 & 7.71 \\ 
2 & 0.559 &    NA & 0.3157 & 9.795 & 295.8 & 1.65 & 9.31 & 1.99 & 7.71 \\ 
\midrule
& \it $\mu$: & 1.395 & 0.3168 & 9.795 & 295.7 & 1.641 & 9.29 & 1.98 & 7.73 \\ 
& $\sigma_\mu$: & 0.004 & 0.0004 & 0.003 &  0.1 & 0.005 & 0.01 & 0.01 & 0.05 \\ 
\midrule 
\multicolumn{10}{c}{\emph{Cape Grim Air Archive tank: 1.12.2004 (UAN: 997089)}} \\ 
1 & 0.565 &    NA &     NA & 9.801 & 295.5 & 1.65 & 9.33 & 1.97 & 7.66 \\ 
1 & 0.300 &    NA &     NA & 9.798 & 295.4 & 1.64 & 9.34 & 1.91 & 7.77 \\ 
1 & 0.137 &    NA &     NA & 9.791 & 295.7 & 1.65 & 9.34 & 1.98 & 7.72 \\ 
1 & 0.074 &    NA &     NA & 9.779 & 295.5 & 1.66 & 9.39 & 2.02 & 7.89 \\ 
1 & 0.611 &    NA &     NA & 9.796 & 295.4 & 1.65 & 9.25 & 1.98 & 7.53 \\ 
1 & 0.598 &    NA &     NA & 9.802 & 295.4 & 1.63 & 9.21 & 1.94 & 7.56 \\ 
1 & 0.575 &    NA &     NA & 9.803 & 295.2 & 1.65 & 9.25 & 1.97 & 7.52 \\ 
2 & 0.610 & 1.389 & 0.3209 & 9.795 & 295.4 & 1.65 &   NA & 1.97 & 7.74 \\ 
2 & 0.590 & 1.388 & 0.3184 & 9.792 & 295.5 & 1.64 &   NA & 1.98 & 8.05 \\ 
3 & 0.493 & 1.382 &     NA & 9.798 & 295.6 & 1.64 & 9.38 & 1.95 & 7.92 \\ 
3 & 0.481 & 1.381 & 0.3216 & 9.789 & 295.2 & 1.65 &   NA & 1.99 & 7.72 \\ 
3 & 0.181 & 1.407 &     NA &    NA & 295.4 &   NA &   NA &   NA &   NA \\ 
3 & 0.177 & 1.397 & 0.3179 & 9.797 & 295.9 & 1.65 &   NA & 2.02 & 7.53 \\ 
3 & 0.170 & 1.382 & 0.3169 & 9.779 & 295.0 & 1.65 &   NA & 1.94 & 7.76 \\ 
3 & 0.163 & 1.404 & 0.3160 & 9.818 & 294.7 & 1.64 &   NA & 1.95 & 7.64 \\ 
3 & 0.064 & 1.382 & 0.3182 &    NA &   NA & 1.66 & 9.37 & 2.01 & 7.94 \\ 
3 & 0.233 & 1.393 & 0.3190 & 9.765 & 295.7 & 1.64 & 9.36 & 2.02 & 7.90 \\ 
\midrule
& \it $\mu$: & 1.391 & 0.3186 & 9.794 & 295.4 & 1.647 & 9.32 & 1.97 & 7.74 \\ 
& $\sigma_\mu$: & 0.003 & 0.0007 & 0.003 &  0.1 & 0.002 & 0.02 & 0.01 & 0.04 \\ 
\midrule 
\multicolumn{10}{c}{\emph{Cape Grim Air Archive tank: 16.12.2010 (UAN: 20101520)}} \\ 
1 & 0.634 &    NA &     NA & 9.802 & 295.5 & 1.65 & 9.29 & 1.96 & 7.68 \\ 
1 & 0.317 &    NA &     NA & 9.805 & 295.3 & 1.65 & 9.26 & 1.96 & 7.92 \\ 
1 & 0.148 &    NA &     NA & 9.800 & 295.7 & 1.65 & 9.37 & 2.00 & 7.59 \\ 
1 & 0.066 &    NA &     NA & 9.807 & 297.0 & 1.65 & 9.39 & 1.99 & 7.88 \\ 
1 & 0.626 &    NA &     NA & 9.802 & 295.6 & 1.64 & 9.23 & 1.98 & 7.46 \\ 
1 & 0.342 &    NA &     NA & 9.802 & 295.2 & 1.64 & 9.33 & 1.98 & 7.54 \\ 
1 & 0.171 &    NA &     NA & 9.793 & 295.7 & 1.66 & 9.36 & 1.95 & 7.46 \\ 
1 & 0.082 &    NA &     NA & 9.792 & 295.4 & 1.66 & 9.34 & 1.97 & 7.67 \\ 
1 & 0.044 &    NA &     NA &    NA & 296.5 & 1.68 & 9.46 & 2.01 & 7.85 \\ 
\midrule
& \it $\mu$: &    NA &     NA & 9.800 & 295.8 & 1.652 & 9.34 & 1.98 & 7.67 \\ 
& $\sigma_\mu$: &    NA &     NA & 0.002 &  0.2 & 0.004 & 0.02 & 0.01 & 0.06 \\ 
\midrule 
\multicolumn{10}{c}{\emph{Cape Grim Air Archive tank: 4.5.2011 (UAN: 20110462)}} \\ 
1 & 0.606 & 1.401 & 0.3186 & 9.789 & 295.4 & 1.65 & 9.36 & 2.02 & 7.82 \\ 
1 & 0.605 & 1.387 & 0.3163 & 9.799 & 295.2 & 1.63 & 9.28 & 1.99 & 7.69 \\ 
1 & 0.590 & 1.383 & 0.3163 & 9.800 & 295.4 & 1.63 & 9.28 & 2.00 & 7.65 \\ 
1 & 0.567 & 1.383 & 0.3174 & 9.798 & 295.8 & 1.62 & 9.30 & 2.00 & 7.69 \\ 
2 & 0.623 & 1.383 & 0.3171 & 9.798 & 295.5 & 1.65 & 9.28 &   NA &   NA \\ 
2 & 0.604 & 1.387 & 0.3155 & 9.808 & 295.2 & 1.65 & 9.29 & 1.97 &   NA \\ 
2 & 0.580 & 1.391 & 0.3171 & 9.794 & 295.9 & 1.66 & 9.34 & 1.97 & 7.72 \\ 
2 & 0.315 & 1.387 & 0.3145 & 9.814 & 296.0 & 1.62 &   NA & 1.97 & 7.52 \\ 
2 & 0.301 & 1.408 & 0.3158 & 9.789 & 295.4 & 1.65 & 9.33 & 2.00 & 7.77 \\ 
\midrule
& \it $\mu$: & 1.390 & 0.3165 & 9.799 & 295.5 & 1.640 & 9.31 & 1.99 & 7.70 \\ 
& $\sigma_\mu$: & 0.003 & 0.0004 & 0.003 &  0.1 & 0.005 & 0.01 & 0.01 & 0.04 \\ 
\bottomrule
\end{longtable}
}

\end{document}